\documentclass[sigconf,screen]{acmart}

\setcopyright{cc}
\setcctype{by}
\acmDOI{10.1145/3832783.3837542}
\acmYear{2026}
\copyrightyear{2026}
\acmISBN{979-8-4007-2882-2/2026/10}
\acmConference[ASE '26]{Proceedings of the 41st IEEE/ACM International Conference on Automated Software Engineering}{October 12--16, 2026}{Munich, Germany}
\acmBooktitle{Proceedings of the 41st IEEE/ACM International Conference on Automated Software Engineering (ASE '26), October 12--16, 2026, Munich, Germany}
\acmSubmissionID{ase26main-p2788-p}
\received{2026-03-27}
\received[accepted]{2026-06-18}

\usepackage{microtype}
\usepackage{multirow}
\usepackage{algorithm}
\usepackage{algpseudocode}
\usepackage{pifont}
\usepackage{enumitem}
\usepackage{float}
\usepackage{afterpage}
\usepackage{listings}
\usepackage{xcolor}
\usepackage{newfloat}
\usepackage{tcolorbox}

\DeclareFloatingEnvironment[fileext=lol,name=Listing]{listing}

\definecolor{addgreen}{RGB}{34,139,34}
\definecolor{delred}{RGB}{178,34,34}
\definecolor{addbg}{RGB}{230,255,230}
\definecolor{delbg}{RGB}{255,230,230}
\definecolor{highlight}{RGB}{255,215,0}
\definecolor{structblue}{RGB}{0,102,204}

\tcbuselibrary{skins}
\newtcolorbox{findingbox}[1][]{%
  enhanced,
  colback=gray!8,
  colframe=gray!60,
  boxrule=0.5pt,
  arc=3pt,
  left=6pt, right=6pt, top=4pt, bottom=4pt,
  fonttitle=\bfseries,
  #1
}
\newcounter{findingctr}

\newcommand{\finding}[2]{%
  \refstepcounter{findingctr}%
  \label{#1}%
  \begin{findingbox}[left=5pt,right=5pt,top=3pt,bottom=3pt]
  \textbf{Finding \thefindingctr.} #2
  \end{findingbox}
}

\newcounter{rqctr}
\newcommand{\rqref}[1]{RQ\ref{#1}}
\newcommand{\rqrange}[2]{\rqref{#1}--\rqref{#2}}
\newcommand{\rqparen}[1]{(\rqref{#1})}
\newcommand{\rqitem}[3]{\item \textbf{\rqref{#1} (#2)}: #3}
\newcommand{\rqsubsection}[3]{%
  \refstepcounter{rqctr}%
  \label{#1}%
  \subsection{RQ\therqctr: #2}%
  \label{#3}%
}

\title{Benchmarking Automated Security Patch Backporting: How Far Are We?}

\author{Jincheng Yang}
\orcid{0009-0002-2859-7921}
\affiliation{%
  \institution{Xidian University}
  \department{School of Cyber Engineering}
  \city{Xi'an}
  \state{Shaanxi}
  \country{China}
}
\email{yangjc@stu.xidian.edu.cn}

\author{Yulong Fu}
\correspondingauthor
\authornote{Yulong Fu and Chengwei Liu are the corresponding authors.}
\orcid{0000-0002-2058-3405}
\affiliation{%
  \institution{Xidian University}
  \department{School of Cyber Engineering and State Key Laboratory of ISN}
  \city{Xi'an}
  \state{Shaanxi}
  \country{China}
}
\email{ylfu@xidian.edu.cn}

\author{Chengwei Liu}
\correspondingauthor
\authornotemark[1]
\orcid{0000-0003-1175-2753}
\affiliation{%
  \institution{Nankai University}
  \department{College of Cryptology and Cyber Science}
  \city{Tianjin}
  \country{China}
}
\email{chengwei.liu@nankai.edu.cn}

\author{Lyuye Zhang}
\orcid{0000-0003-3087-9645}
\affiliation{%
  \institution{Nankai University}
  \department{College of Cryptology and Cyber Science}
  \city{Tianjin}
  \country{China}
}
\affiliation{%
  \institution{Nanyang Technological University}
  \city{Singapore}
  \country{Singapore}
}
\email{zh0004ye@e.ntu.edu.sg}

\author{Fangyuan Zhang}
\orcid{0009-0000-9599-1369}
\affiliation{%
  \institution{Nankai University}
  \department{College of Cryptology and Cyber Science}
  \city{Tianjin}
  \country{China}
}
\email{fangyuanzhang@mail.nankai.edu.cn}

\author{Bingyang Ren}
\orcid{0009-0008-3362-8837}
\affiliation{%
  \institution{Xidian University}
  \department{School of Cyber Engineering}
  \city{Xi'an}
  \state{Shaanxi}
  \country{China}
}
\email{24151213637@stu.xidian.edu.cn}

\author{Yang Liu}
\orcid{0000-0001-7300-9215}
\affiliation{%
  \institution{Nanyang Technological University}
  \city{Singapore}
  \country{Singapore}
}
\email{yangliu@ntu.edu.sg}

\author{Hui Li}
\orcid{0000-0001-8310-7169}
\affiliation{%
  \institution{Xidian University}
  \department{School of Cyber Engineering and State Key Laboratory of ISN}
  \city{Xi'an}
  \state{Shaanxi}
  \country{China}
}
\email{lihui@mail.xidian.edu.cn}

\renewcommand{\shortauthors}{J. Yang, Y. Fu, C. Liu, L. Zhang, F. Zhang, B. Ren, Y. Liu, and H. Li}

\begin{document}
\raggedbottom
\begin{abstract}
Automated security patch backporting is critical for mitigating N-day vulnerabilities.
Recent tools report success rates above 80\% on their respective datasets.
However, these evaluations are often confined to homogeneous environments, such as one repository or specific project versions.
Consequently, it remains unclear how well these tools generalize beyond their originally targeted scenarios.

We present \textsf{Porting Benchmark}, a curated dataset of 1{,}234 security patch backporting cases spanning cross-version, cross-branch, and cross-repository scenarios, paired with a common evaluation framework.
Using this benchmark, we evaluate five tools spanning program analysis, LLM prompting, and LLM agents under aligned settings.
Our results show that aligned evaluation changes the apparent performance landscape: PortGPT and TSBPort remain comparatively strong on the Replication Dataset, while FixMorph and Mystique degrade substantially under the common protocol.
Performance degrades sharply on structurally complex patches: the best commit-level success rate falls from 85.2\% on Type-I patches to 24.0\% on Type-IV.
We identify four root-cause categories (missing target API awareness, cross-version semantic mismatch, non-local dependency propagation failure, and patch construction or localization failure) and derive concrete directions for next-generation tool design.
On a 45-case dynamically validated subset with verified test cases and constructed POCs, we further observe that reference-based benchmark scores do not fully capture real-world remediation: exact match sharply under-credits harder target adaptations, while executable validation reveals residual integration failures in the target that static reference agreement misses.
Executable-feedback refinement provides limited but measurable recovery on the hardest executable cases.
\end{abstract}



\maketitle

\section{Introduction}
\label{sec:intro}

Automated security patch backporting is critical for mitigating N-day vulnerabilities~\cite{vpsurvey2024}.
When a vulnerability is patched in one version, the fix must propagate to other maintained versions, long-term support (LTS) branches, or downstream forks. This process is known as patch backporting~\cite{linuxporting2024}.
The Linux kernel alone maintains dozens of active stable branches simultaneously, and empirical studies have documented significant delays and inconsistencies in how patches propagate across this ecosystem~\cite{vulngap2025}.
Manual backporting is time-consuming, error-prone, and demands deep understanding of both the vulnerability and the target codebase, motivating a growing line of research on automated tools.

Automated patch backporting tools span three paradigms. Traditional program-analysis methods include FixMorph~\cite{fixmorph2021}, TSBPort~\cite{tsbport2023}, and PatchWeave~\cite{patchweave2021}; LLM-based prompting includes Mystique~\cite{mystique2024}, PPatHF~\cite{ppathf2024}, and MigGPT~\cite{miggpt2025}; and PortGPT~\cite{portgpt2025} uses an LLM agent.
These tools have reported promising results on their respective datasets, with success rates ranging from 42\% to 95\%.

However, these results are not directly comparable.
Each tool uses its own dataset, evaluation setting, success criterion, and input assumption.
For instance, FixMorph uses compilation-filtered equivalence analysis, TSBPort uses PDG matching, Mystique uses edit distance, and PPatHF reports syntactic equivalence together with AED/RED-style edit-distance metrics. A tool reporting 80\% compilation success and another reporting an edit-similarity score cannot be meaningfully compared.
Moreover, each tool has naturally been evaluated on its originally targeted scenario: FixMorph and TSBPort on Linux cross-version data, Mystique on same-repository branches, PPatHF on a single fork pair (Vim$\to$Neovim).
Whether the reported performance generalizes to other backporting contexts (e.g., cross-repository backporting, long multi-file patches, or structurally divergent codebases) remains an open question.
Finally, even under a unified benchmark, it remains unclear whether a single static metric such as reference-based agreement adequately captures whether the generated patch actually blocks exploitation and preserves intended behavior under execution.

These observations expose four gaps in our understanding of automated patch backporting. \textbf{Comparability gap.} Incompatible metrics and input assumptions prevent fair cross-tool comparison. \textbf{Generalization gap.} Evaluation within originally targeted scenarios leaves capability boundaries unknown. \textbf{Coverage gap.} Datasets concentrated in specific projects or narrow settings may inflate reported performance. \textbf{Failure-explanation gap.} Without systematic failure analysis, why tools fail and which capabilities are missing remain unclear.

In this paper, we address these gaps through \textsf{Porting Benchmark}, a comprehensive benchmark designed for systematic cross-tool evaluation of patch backporting tools.
We integrate and curate existing public datasets into a unified format, construct new evaluation data from recent vulnerability sources (post-2023) to minimize LLM data leakage, define evaluation criteria across multiple dimensions (cross scenario, porting type, patch scale, and information level), and evaluate five state-of-the-art tools under aligned settings.

\begin{listing}[t]
  \centering
  \begin{tcolorbox}[
    enhanced,
    colback=white,
    colframe=gray!45,
    boxrule=0.4pt,
    arc=1pt,
    left=3pt,right=3pt,top=3pt,bottom=3pt,
    boxsep=0pt
  ]
  \begin{minipage}[t]{0.485\linewidth}
  \centering
  \textbf{(a) Source: \texttt{torvalds/linux}}\par\smallskip
  \begin{lstlisting}[language=C,basicstyle=\ttfamily\scriptsize,escapeinside={(*@}{@*)},numbers=none,frame=none,xleftmargin=0pt,xrightmargin=0pt,aboveskip=0pt,belowskip=0pt]
  (*@\colorbox{addbg}{\color{structblue}\textbf{// \ding{182} Struct extension}}@*)
  struct ext_arg {
    ...
  (*@\colorbox{addbg}{\color{addgreen}+ bool iowait; /* new */}@*)
  };
  
  (*@\colorbox{addbg}{\color{structblue}\textbf{// \ding{183} Param via struct}}@*)
  int
  __io_cqring_wait_schedule(
    struct io_ring_ctx *ctx,
    struct io_wait_queue *iowq,
  (*@\colorbox{addbg}{\color{addgreen}+ struct ext\_arg *ext\_arg,}@*)
    ...) {
  
  (*@\colorbox{addbg}{\color{structblue}\textbf{// \ding{184} Struct field access}}@*)
  (*@\colorbox{delbg}{\color{delred}- if (current\_pending\_io())}@*)
  (*@\colorbox{addbg}{\color{addgreen}+ if (ext\_arg->iowait \&\&}@*)
  (*@\colorbox{addbg}{\color{addgreen}+     current\_pending\_io())}@*)
      current->in_iowait = 1;
  }
  \end{lstlisting}
  \end{minipage}
  \hfill
  \begin{minipage}[t]{0.485\linewidth}
  \centering
  \textbf{(b) Target: \texttt{GrapheneOS/kernel}}\par\smallskip
  \begin{lstlisting}[language=C,basicstyle=\ttfamily\scriptsize,escapeinside={(*@}{@*)},numbers=none,frame=none,xleftmargin=0pt,xrightmargin=0pt,aboveskip=0pt,belowskip=0pt]
  (*@\colorbox{highlight}{\color{structblue}\textbf{// \ding{182} No ext\_arg struct}}@*)
  /* Older target has no
     ext_arg abstraction. */
    ...
  
  (*@\colorbox{highlight}{\color{structblue}\textbf{// \ding{183} Direct bool param}}@*)
  int
  io_cqring_wait_schedule(
    struct io_ring_ctx *ctx,
    struct io_wait_queue *iowq,
  (*@\colorbox{addbg}{\color{addgreen}+ bool iowait) \{}@*)
    ...
  
  (*@\colorbox{highlight}{\color{structblue}\textbf{// \ding{184} Direct variable}}@*)
  (*@\colorbox{delbg}{\color{delred}- if (current\_pending\_io())}@*)
  (*@\colorbox{addbg}{\color{addgreen}+ if (iowait \&\&}@*)
  (*@\colorbox{addbg}{\color{addgreen}+     current\_pending\_io())}@*)
      current->in_iowait = 1;
  }
  \end{lstlisting}
  \end{minipage}
  \end{tcolorbox}
  \caption{Motivating example of cross-repository backporting from Linux to GrapheneOS (\texttt{IORING\_ENTER\_NO\_IOWAIT})}
  \label{lst:porting-example}
\end{listing}

Our evaluation shows that aligned evaluation changes the apparent performance landscape: PortGPT and TSBPort remain comparatively strong on the Replication Dataset, whereas FixMorph and Mystique degrade substantially under the common protocol \rqparen{rq:stability}. Structural complexity is the dominant bottleneck, with the best commit-level success falling from 85.2\% on Type-I patches to 24.0\% on Type-IV \rqparen{rq:scenario}. On the 45-case executable subset, PortGPT reaches 80.0\% for both S-Succ and Full validation but exhibits two case-level mismatches \rqparen{rq:gap}. Root-cause analysis identifies patch construction or localization failure and non-local dependency propagation failure as the two most frequent categories \rqparen{rq:failure}, while executable-feedback refinement provides limited but measurable recovery on eligible hard cases \rqparen{rq:improve}.

This paper makes the following contributions:
\begin{enumerate}[leftmargin=*]
    \item We construct and publicly release \textsf{Porting Benchmark}, a curated dataset of 1{,}234 patch backporting cases (600 for replication and 634 for evaluation under the common static scope that excludes tests and non-target files) spanning cross-version, cross-branch, and cross-repository scenarios, with rich metadata and a reproducible evaluation methodology.
    \item We conduct the first comprehensive comparison across five state-of-the-art patch backporting tools under aligned evaluation settings, exposing significant performance gaps and previously unknown generalization boundaries.
    \item We augment 45 cases from the Evaluation Dataset with verified test cases and constructed POC scripts, and evaluate four patch backporting tools and SWE-agent (GPT-4o) under a common file-level input setting.
    \item We provide a root cause analysis that identifies four recurring capability deficits: missing target API awareness, cross-version semantic mismatch, non-local dependency propagation failure, and patch construction or localization failure.
    \item We conduct a targeted executable-feedback refinement study on 29 hard PortGPT cases (Type-III/IV or cross-version) among the 45 cases with executable validation assets.
\end{enumerate}

\begin{table*}[!htbp]
  \centering
  \caption{Patch backporting tool pipeline: stages and strategies (\ding{51}=used, $^\dagger$=excluded)}
  \label{tab:tool-pipeline}
  \footnotesize
  \setlength{\tabcolsep}{2pt}
  \renewcommand{\arraystretch}{1.08}
  \newcommand{\cmark}{\ding{51}}%
  \begin{tabular}{@{}p{1.4cm}p{1.05cm}p{0.7cm}p{6.5cm}|cccccccc@{}}
  \toprule
  \textbf{Stage} & \textbf{Gran.} & \textbf{ID} & \textbf{Strategy} &
  \rotatebox{70}{\textbf{FixMorph}~\cite{fixmorph2021}} &
  \rotatebox{70}{\textbf{TSBPort}~\cite{tsbport2023}} &
  \rotatebox{70}{\textbf{PatchWeave$^\dagger$}~\cite{patchweave2021}} &
  \rotatebox{70}{\textbf{SKYPORT$^\dagger$}~\cite{skyport2022}} &
  \rotatebox{70}{\textbf{Mystique}~\cite{mystique2024}} &
  \rotatebox{70}{\textbf{PPatHF}~\cite{ppathf2024}} &
  \rotatebox{70}{\textbf{MigGPT$^\dagger$}~\cite{miggpt2025}} &
  \rotatebox{70}{\textbf{PortGPT}~\cite{portgpt2025}} \\
  \midrule
  \multirow{6}{*}{\shortstack[l]{Stage 1:\\Localization}}
   & Repo      & S1-1: & File-level path matching                    & \cmark & \cmark &        &        &        &        &        & \cmark \\
   & Function  & S1-2: & AST/PDG-based function mapping              & \cmark & \cmark &        &        &        &        &        &        \\
   & Function  & S1-3: & LLM-based function identification            &        &        &        &        & \cmark & \cmark &        & \cmark \\
   & Statement & S1-4: & Hunk-level offset matching                   & \cmark & \cmark &        &        &        &        &        &        \\
   & Statement & S1-5: & LLM-based hunk/line localization             &        &        &        &        &        &        &        & \cmark \\
   & Function  & S1-6: & Code fingerprint-based migration point identification &        &        &        &        &        &        & \cmark &        \\
  \midrule
  \multirow{7}{*}{\shortstack[l]{Stage 2:\\Transform.}}
   & Statement & S2-1: & AST edit script (anti-unification)           & \cmark &        &        &        &        &        &        &        \\
   & Statement & S2-2: & PDG-guided semantic transplant               &        & \cmark &        &        &        &        &        &        \\
   & Function  & S2-3: & Single-round LLM prompting                   &        &        &        &        & \cmark & \cmark & \cmark &        \\
   & Hunk      & S2-4: & Iterative LLM agent with tool use            &        &        &        &        &        &        &        & \cmark \\
   & Function  & S2-5: & Signature-guided context slicing for LLM     &        &        &        &        & \cmark & \cmark & \cmark &        \\
   & Function  & S2-6: & Symbolic-execution-based patch transplantation (requires tests) &        &        & \cmark &        &        &        &        &        \\
   & Statement & S2-7: & Vulnerability-type-specific static analysis backporting &        &        &        & \cmark &        &        &        &        \\
  \midrule
  \multirow{7}{*}{\shortstack[l]{Stage 3:\\Validation}}
   & Binary     & S3-1: & Compilation success                          & \cmark &        &        &        &        &        &        &        \\
   & Structural & S3-2: & PDG equivalence matching                     &        & \cmark &        &        &        &        &        &        \\
   & Textual    & S3-3: & Edit distance / diff similarity              &        &        &        &        & \cmark &        &        &        \\
   & Semantic   & S3-4: & Syn.\ Eq.\ + AED/RED-style metrics           &        &        &        &        &        & \cmark &        &        \\
   & Manual     & S3-5: & Human expert review                          &        &        &        &        &        &        &        & \cmark \\
   & Dynamic    & S3-6: & Test-suite validation                         &        &        & \cmark &        &        &        &        &        \\
   & Textual    & S3-7: & Completion rate against reference patches     &        &        &        &        &        &        & \cmark &        \\
  \bottomrule
  \end{tabular}
\end{table*}

\section{Background}
\label{sec:background}

\subsection{Patch Backporting}
\label{subsec:challenges}

Security patch backporting adapts a known-correct fix from one code context to another whose surrounding implementation may have diverged substantially.
Unlike a direct cherry-pick, the target codebase may require the fix to be relocated, rewritten, or revalidated under different interfaces and control-flow structures.
These differences commonly include (1) syntactic drift such as variable or function renaming, (2) structural drift such as changed control flow or refactored code blocks, (3) API evolution where function signatures or parameters differ, and (4) context dependencies where surrounding code affects how the patch should be applied.
Viewed end-to-end, patch backporting involves three coupled stages: Localization (finding where the fix should be applied), Transformation (adapting the fix to the target context), and Validation (assessing whether the adapted patch preserves the fix intent).

\subsection{Existing Tools}
\label{subsec:existing-tools}

Existing patch backporting tools span traditional program analysis, LLM-based prompting, and LLM agents.
Despite this variety, their reported results are difficult to compare directly because they target different backporting scenarios, operate at different execution granularities, assume different amounts of target-location information, and validate outputs with incompatible criteria.
We briefly review representative tools below with emphasis on these evaluation-relevant differences.

\textbf{Traditional Program Analysis.}
FixMorph~\cite{fixmorph2021} was among the first tools for Linux kernel backporting. It employs anti-unification and AST-level transformations. Its original evaluation follows a layered pipeline: compilation serves as an initial filter, followed by equivalence analysis on the resulting subset.
TSBPort~\cite{tsbport2023} introduces semantic-aware backporting based on program dependence graphs (PDGs). It evaluates via PDG matching, which captures structural correctness but requires compilation infrastructure.
Both tools primarily target Linux kernel backporting and operate at statement-level granularity.
PatchWeave~\cite{patchweave2021} uses symbolic execution for cross-program transplantation but requires test cases as input.
SKYPORT~\cite{skyport2022} targets web application injection vulnerabilities via static analysis.

\textbf{LLM-based Prompting.}
Mystique~\cite{mystique2024} combines semantic and syntactic slicing signatures with LLMs. It evaluates via edit distance.
PPatHF~\cite{ppathf2024} addresses hard-fork scenarios using code reduction. It reports syntactic equivalence together with AED/RED-style edit-distance metrics.
MigGPT~\cite{miggpt2025} introduces code fingerprints for out-of-tree Linux patch migration. Its implementation is not publicly available.

\textbf{LLM Agent.}
PortGPT~\cite{portgpt2025} introduces an agentic approach with tool use and iterative workflows. It operates at hunk-level granularity and relies on manual verification.

These differences are methodological rather than cosmetic: a tool reporting compilation success, another reporting similarity-style scores, and another relying on manual review are not operating under a shared notion of success.

\begin{figure}[t]
  \centering
  \includegraphics[width=\columnwidth]{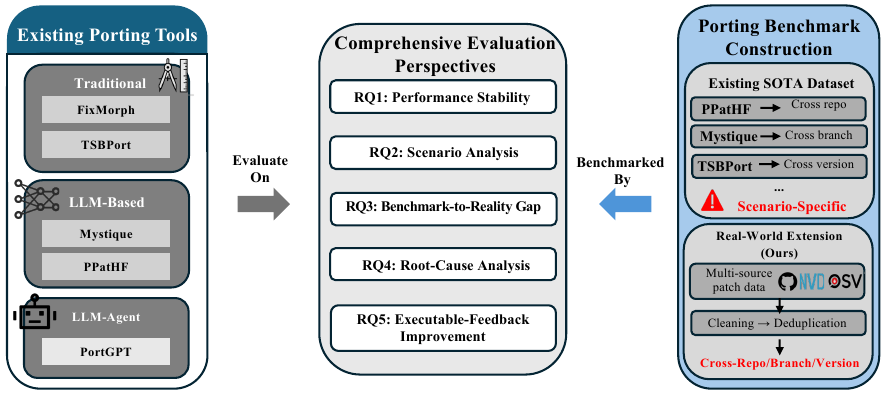}
  \caption{Overview of the unified benchmark evaluation workflow}
  \Description{An overview figure showing the study workflow from benchmark construction to tool evaluation and analysis.}
  \label{fig:overview}
\end{figure}

\textbf{Technical Pipeline Decomposition.}
Figure~\ref{fig:overview} outlines the study workflow, and Table~\ref{tab:tool-pipeline} decomposes tool pipelines into Localization, Transformation, and Validation.
Tools differ in every pipeline stage, not only in their overall paradigm.
Traditional tools emphasize program-analysis-based localization and transformation, whereas LLM-based tools rely more heavily on context selection and generation.
Validation remains entirely tool-specific, which is a key driver of the comparability gap we address.
This decomposition informs our tool selection (Section~\ref{subsec:tool-selection}) and highlights why a common evaluation framework is needed.

\subsection{Evaluation Gaps}
\label{subsec:gaps}

A concrete cross-repository case illustrates why the central empirical questions remain unresolved.
Listing~\ref{lst:porting-example} shows a real cross-repository case.
The source patch extends an existing struct (\ding{182}), passes it through function parameters (\ding{183}), and accesses it via struct field dereference (\ding{184}).
The target codebase lacks this struct entirely.
The implementation must be restructured with direct parameter passing through a 4-function call chain.
Such cases demand non-local structural adaptation rather than simple syntactic transfer.
As in Section~\ref{subsec:rq-failure}, even the best-performing tool fails on this case.

This example also exposes why existing evidence remains incomplete.
Current evaluations vary substantially in scenario coverage, validation criteria, granularity, and dataset composition, making it difficult to tell how far current automation actually generalizes.
We highlight four gaps:

\begin{itemize}[leftmargin=*,noitemsep]
    \item \textbf{Scenario-coverage gap.} Most tools are designed for and validated in a single setting (cross-version, cross-branch, or cross-repository). It is unclear whether their reported effectiveness transfers to other backporting contexts.
    \item \textbf{Validation-criteria gap.} Tools adopt different success criteria (e.g., compilation, similarity scores, PDG equivalence, or manual judgment), so results are difficult to compare even when datasets partially overlap and often stop short of a common end-to-end notion of success.
    \item \textbf{Granularity-assumption gap.} Tools operate at different execution granularities (file, function, or hunk) and rely on different degrees of localization assistance. Without controlling these assumptions, comparisons conflate localization capability with transformation quality.
    \item \textbf{Dataset-representativeness gap.} Public evaluations are dominated by a few ecosystems (e.g., the Linux kernel) or narrow fork scenarios. This can over-represent easy cases and mask failure modes in diverse, structurally divergent patches.
\end{itemize}

Together, these issues prevent the literature from answering a basic empirical question: how far current patch backporting tools generalize under aligned assumptions and common evaluation metrics.
They motivate our study design in Section~\ref{sec:study-design}, where we construct a benchmark with diverse scenarios, controlled localization assumptions, and a common evaluation framework.

\section{Study Design}
\label{sec:study-design}

We address the gaps identified in Section~\ref{sec:background} through a common evaluation methodology covering benchmark dimensions, tool selection, data construction, evaluation metrics, and research questions.

\subsection{Benchmark Dimensions}
\label{subsec:dimensions}

We first establish a unified taxonomy for categorizing patch backporting cases.
We define two orthogonal dimensions following prior work: Cross Scenario (source--target relationship) and Porting Type (complexity).

\subsubsection{Cross Scenario}
We distinguish three source--target relationships:

\begin{itemize}[leftmargin=*,noitemsep]
    \item \textbf{Cross-version}: Backporting between different version lines of the same software (e.g., from Linux kernel v6.1 mainline to v5.15 stable). This is the primary scenario targeted by FixMorph and TSBPort.
    \item \textbf{Cross-branch}: Backporting between branches within the same repository that do not represent distinct version lines (e.g., cherry-picks across development or feature branches). This is the scenario targeted by Mystique and PortGPT.
    \item \textbf{Cross-repository}: Backporting between different repositories, such as forks (e.g., Vim to Neovim) or upstream-to-downstream distributions. This is the scenario targeted by PPatHF.
\end{itemize}

\textbf{Linux Kernel Classification.}
Linux stable/LTS lines use separate Git branches (e.g., \texttt{linux-4.19.y}).
However, they represent distinct release lines with strict backporting rules~\cite{kernelreleases,stable_kernel_rules}.
We therefore classify mainline-to-stable/LTS backports as cross-version rather than cross-branch.

\subsubsection{Porting Type and Path Change}
We characterize backporting complexity along two orthogonal dimensions: code adaptation complexity and path change.

\textbf{Code Adaptation Complexity.}
Following TSBPort~\cite{tsbport2023} and FixMorph~\cite{fixmorph2021}, we classify code adaptation complexity into four types.
Type-I (Identical): the patch can be applied without any code modifications.
Type-II (Location Change): the patch requires only location adjustments (e.g., line numbers) but no code transformation.
Type-III (Syntactic Adaptation): the patch requires syntactic modifications (e.g., renaming variables or functions) to fit the target context.
Type-IV (Structural Change): the patch requires logical or structural modifications (e.g., adding or removing statements) while preserving the fix semantics.
The motivating example in Listing~\ref{lst:porting-example} is a Type-IV case.
The source uses a struct-mediated parameter, but the target requires restructuring to direct parameter passing.

\textbf{Path Change.}
We also track whether the target file path differs from the source.
Prior work typically assumes patches apply to the same file path.
However, real-world backporting often involves file renaming or directory restructuring.
A patch may be Type-I in code adaptation yet still require localization effort due to path change.
We analyze path change as a separate dimension in our evaluation (Section~\ref{subsubsec:eval-scope}).

\subsection{Tool Selection}
\label{subsec:tool-selection}

From the eight tools reviewed in Section~\ref{subsec:existing-tools}, we select five that cover all three paradigms.
Table~\ref{tab:tool-summary} summarizes the selected tools and their original evaluation setups.
We apply three selection criteria: (1) the tool must be publicly available; (2) it must support C/C++ programs; (3) it must handle general patch backporting rather than specific vulnerability types only.
Three tools are excluded (marked $^\dagger$ in Table~\ref{tab:tool-pipeline}).
PatchWeave~\cite{patchweave2021} requires test exploits that are unavailable for most vulnerability patches.
SKYPORT~\cite{skyport2022} targets only web injection vulnerabilities.
MigGPT~\cite{miggpt2025} lacks a public implementation.

\subsection{Benchmark Construction}
\label{subsec:benchmark}

We construct Porting Benchmark to improve cross-tool comparability while covering a representative range of real-world scenarios.

\begin{figure}[t]
  \centering
  \includegraphics[width=\columnwidth]{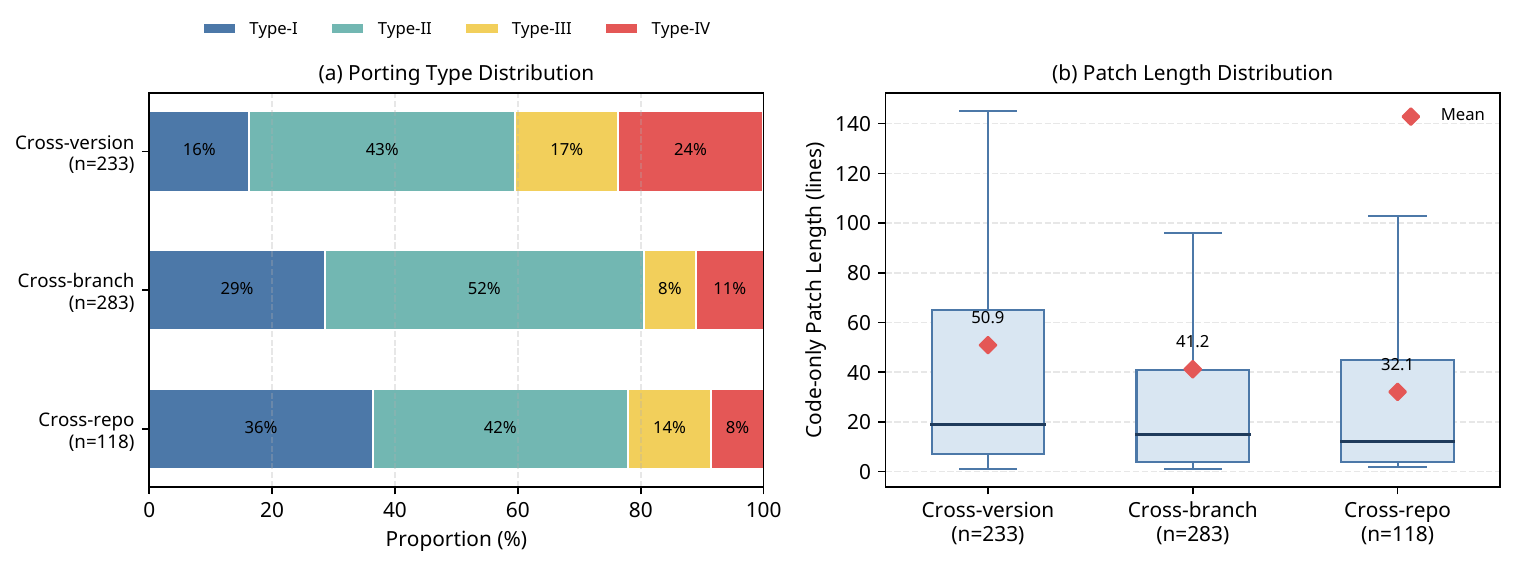}
  \caption{Evaluation Dataset by cross scenario under the common static evaluation scope: (a) porting type distribution; (b) patch length distribution}
  \Description{A two-panel figure showing porting type distribution and patch length distribution across cross-version, cross-branch, and cross-repository scenarios.}
  \label{fig:eval-dataset}
\end{figure}

\subsubsection{Data Collection}
\label{subsubsec:data-sources}

We use a two-tier design: a Replication Dataset for reproducibility-oriented evaluation and an Evaluation Dataset for broader capability assessment.
This split maps directly to our goals: \rqref{rq:stability} isolates comparability effects under aligned metrics, whereas \rqrange{rq:scenario}{rq:failure} probe generalization on newer, scenario-diverse cases.

\textbf{1. Replication Dataset.}
Replication evaluation must respect the constraints of each tool. PPatHF~\cite{ppathf2024} and Mystique~\cite{mystique2024} use LLM prompting and therefore require temporal controls to mitigate leakage, whereas the released FixMorph artifact provides validated build and analysis configurations only for the Linux kernel~\cite{fixmorph2021}. TSBPort~\cite{tsbport2023} and PortGPT~\cite{portgpt2025} can in principle use the full published pool.
We therefore merge the five released datasets, normalize records, and remove duplicates to form a phase-1 pool of 3{,}340 valid commit-level records.
Because this pool is imbalanced across cross scenarios and the prompting tools must be evaluated according to model training dates, we stratify by cross scenario (cross-version vs.\ cross-branch) and era (before vs.\ after July~1, 2022). Mystique's later cutoff of August~23, 2023 is used only to derive its subset.
Following standard references on stratified sampling and disproportionate allocation~\cite{neyman1934,thompson2012}, we then draw a disproportionate 600-case sample with fixed seed 42, balancing cross-version and cross-branch while emphasizing post-cutoff cases.
The resulting replication tier expands to 917 file-level pairs and 2{,}020 functions.
A separate \rqref{rq:stability} subset is then derived for each tool from this shared pool: 367~cases for PPatHF, 224 for Mystique, 285 for FixMorph, and 600 each for TSBPort and PortGPT.
This is why \rqref{rq:stability} reports per-tool denominators rather than a single shared $n$: forcing one denominator would either violate tool constraints or discard valid replication data.

\begin{table}[t]
  \centering
  \caption{Selected tools for evaluation: original setups}
  \label{tab:tool-summary}
  \resizebox{\columnwidth}{!}{%
  \begin{tabular}{@{}l|lllll@{}}
  \toprule
  \textbf{Tool} & \textbf{\shortstack{Exec.\\Gran.}} & \textbf{Target Scenario} & \textbf{Data Source} & \textbf{Metric} & \textbf{Scale} \\
  \midrule
  FixMorph & Statement & Linux mainline$\to$stable & Linux kernel & Layered eval. & 350 \\
  TSBPort & Statement & Linux mainline$\to$stable & NVD + Linux & PDG match & 1,815 \\
  Mystique & Function & Same-repo branches & 33 repositories & Edit dist. & 1,359 \\
  PPatHF & Function & Vim$\to$Neovim fork & Vim/Neovim & Syn.\ Eq./AED/RED & 310 \\
  PortGPT & Hunk & Same-repo branches & 6 repositories & Manual & 537 \\
  \bottomrule
  \end{tabular}%
  }
\end{table}

\textbf{2. Evaluation Dataset.}
We build a separate Evaluation Dataset to assess generalization beyond the data used in the original tool papers.
To reduce leakage for LLM-based tools, we collect only 2024+ cases, well after the training cutoffs of the models used by the evaluated systems~\cite{starcoder2023,codellama2023,gpt4o2024}.
No single source covers all scenarios reliably, so we use two complementary collection routes.
For cross-version and cross-branch cases, we collect vulnerability-linked commits from NVD~\cite{nvd}, GitHub Security Advisories~\cite{ghsa}, and OSV~\cite{osv}, then expand candidate pairs with Mystique-style CVE-equivalent matching logic.
For cross-repository cases, we mine explicit downstream backport/cherry-pick commits via the GitHub Search Commits API~\cite{ghsearchapi} and verify the referenced upstream commits.
Under the common static evaluation scope used in \rqrange{rq:scenario}{rq:failure}, which excludes tests, documentation, configuration files, and other non-target files, this process yields 634 evaluation commits spanning 192 CVEs and 80 target repositories (233 cross-version, 283 cross-branch, and 118 cross-repository), expanding to 1{,}299 files and 2{,}515 functions. The artifact records per-case metadata and, for the 45-case executable subset (18 CVEs and 11 target repositories), executable assets and validation commands.

\subsubsection{Ground-Truth Validation}
For each collected record, we extract source and target diffs via \texttt{git show -M}, treat the target diff as ground truth, and parse C/C++ source files with tree-sitter~\cite{treesitter}. We map each diff hunk, via interval intersection, to every function whose source span overlaps the hunk; hunks outside function bodies are retained under their enclosing file. This establishes a hierarchical commit$\to$file$\to$function representation in which hunks remain the atomic changed regions: no hunk is discarded by the mapping. Each hunk contributes to evaluation through its overlapping function(s), or through file-level comparison for non-function hunks. The hierarchy is used only as a unified aggregation coordinate system across tools with different native granularities (hunk, function, or file); it does not alter what any tool consumes at execution time. Hunk boundaries themselves vary with the diff algorithm and context size, so they are a fragile shared denominator; aggregating through overlapping functions or files keeps every edit while remaining comparable across tools. We preserve repository-, file-, and function-level context to support the I0--I2 settings studied later, and annotate each case with cross scenario, porting type, patch scale, and path change. To validate the reliability of these manual labels, two authors with over five years of software-security experience independently annotated cross-scenario and porting-type labels for a stratified random sample of 100~cases from the Evaluation Dataset. Agreement was high (Cohen's $\kappa=1.000$ for cross-scenario and 0.972 for porting type), and the few disagreements were resolved by a third author.
Figure~\ref{fig:eval-dataset} summarizes the main characteristics of the Evaluation Dataset: cross-version cases contain a larger share of Type-III/IV patches and generally larger edits, foreshadowing the harder results in \rqref{rq:scenario}.

\begin{algorithm}[t]
\caption{Five-Level Hierarchical Patch Matching}
\label{alg:validation}
\small
\begin{algorithmic}[1]
\Require Pre-fix target code $B$, tool output code $C_{tool}$, reference code $C_{ref}$, tolerances $\delta$ and $\tau$
\Ensure Exact Match $EM$, Semantic Success $S$-Succ
\State $B' \gets$ \Call{Normalize}{$B$}; $C'_{tool} \gets$ \Call{Normalize}{$C_{tool}$}; $C'_{ref} \gets$ \Call{Normalize}{$C_{ref}$}
\Statex \hrulefill\ \textbf{L1: Identity (Exact Match)}\ \hrulefill
\State $L1 \gets (C'_{tool} \equiv C'_{ref})$
\State $EM \gets L1$
\Statex
\Statex \hrulefill\ \textbf{L2--L5: Segment/Function/Precision/Completeness}\ \hrulefill
\State $D_{tool} \gets$ \Call{Diff}{$B',C'_{tool}$}; $D_{ref} \gets$ \Call{Diff}{$B',C'_{ref}$}
\State $Seg_{tool} \gets$ \Call{ExtractSegments}{$D_{tool}$}
\State $Seg_{ref} \gets$ \Call{ExtractSegments}{$D_{ref}$}
\State $L2 \gets$ \Call{SegmentMatch}{$Seg_{tool},Seg_{ref},\delta,\tau$} \Comment{one-to-one; normalized similarity}
\State $F_{tool} \gets$ \Call{GetTargetFunctions}{$D_{tool}$}; $F_{ref} \gets$ \Call{GetTargetFunctions}{$D_{ref}$}
\State $L3 \gets$ \Call{FunctionOverlap}{$F_{tool},F_{ref},\delta$}
\State $Pos_{tool} \gets$ \Call{ModifiedPositions}{$D_{tool}$}; $Pos_{ref} \gets$ \Call{ModifiedPositions}{$D_{ref}$}
\State $L4 \gets$ \Call{Precision}{$Pos_{tool},Pos_{ref},\delta$} \Comment{no spurious modifications}
\State $L5 \gets$ \Call{Completeness}{$Pos_{tool},Pos_{ref},\delta$} \Comment{no missing modifications}
\State $S$-Succ $\gets EM$
\If{$\neg EM \land L2 \land L3 \land L4 \land L5$}
    \State $S$-Succ $\gets$ \Call{ManualVerify}{$B,C_{tool},C_{ref},D_{tool},D_{ref}$}
\EndIf
\State \Return $EM, S$-Succ
\end{algorithmic}
\end{algorithm}

\subsection{Evaluation Framework}

Because the selected tools use incompatible native metrics, we align their released outputs to common denominators, aggregation levels, and validation metrics as closely as available evidence permits, improving cross-tool comparability.

\subsubsection{Validation Hierarchy}

We adopt a five-level validation hierarchy inspired by TSBPort~\cite{tsbport2023}.
Each level checks one aspect of agreement with the reference code after patching.
Each benchmark case contains the target code immediately before the security fix and the known patch that developers used to fix the target version.
We call this vulnerable target version the \emph{pre-fix target}.
An evaluated tool produces an adapted fix for the pre-fix target; applying this output gives the \emph{tool output code}.
Applying the known target patch to a separate copy of the same pre-fix target produces the \emph{reference code after patching}, which serves as the ground truth.
We normalize both versions by removing comments and standardizing whitespace.
\textbf{L1} requires the normalized tool output code and reference code to be identical.
For non-identical cases, we compute both diffs against the pre-fix target and check structural consistency with tolerance \(\delta=3\) lines.
\textbf{L2} requires every added or deleted block produced by the tool to be paired one-to-one with a distinct block in the reference patch.
A pair is accepted only when both blocks modify the same file, represent the same operation (addition or deletion), their starting lines differ by at most \(\delta\), and their normalized contents satisfy the fixed similarity threshold \(\tau=0.8\).
The similarity is computed after the same comment removal and whitespace normalization applied above.
\textbf{L3} requires each paired block to affect the same target function.
\textbf{L4} enforces precision: every tool-modified position must match a ground-truth modified position.
\textbf{L5} enforces completeness: every ground-truth modified position must be covered by the tool output.
We apply exactly these rules to every evaluated tool: all outputs undergo the same normalization, and L2 uses the same \(\delta\) and pairing conditions.
Algorithm~\ref{alg:validation} summarizes the evaluation procedure.

Because native metrics are incompatible and executable evidence is unavailable for most cases, we report EM and S-Succ as common reference-based assessments.
EM (L1) checks normalized identity and cannot recognize non-exact adaptations.
For non-exact outputs, Semantic Success (S-Succ) requires L2--L5 structural consistency followed by ManualVerify.
S-Succ measures reference consistency, not executable correctness; PPR, TPR, and Full provide separate executable evidence where available.
For tools that emit patches, functions, or files, we reconstruct the corresponding tool output code before evaluation; missing outputs, application failures, and extraction failures remain in the denominator and count as failures.

\paragraph{Manual Verification Protocol.}
ManualVerify reviews only non-exact function-level items that pass L2--L5; functions failing any L2--L5 check are not manually promoted.
Grounded in security-patch validation principles and evidence of incomplete or regressive fixes~\cite{li2025sok,li2017securitypatches}, reviewers apply seven checks: \textbf{MV1 Syntax} flags malformed or corrupted code; \textbf{MV2 Safety/Cleanup} checks guard ordering, resource release, and error paths; \textbf{MV3 Missing/Weakened Fix} checks whether security-relevant reference edits are absent, misplaced, or weakened; \textbf{MV4 Spurious Behavior} rejects unrelated behavior changes; \textbf{MV5 Target Adaptation} accepts target-version API, macro, field, or helper changes that preserve the fix intent; \textbf{MV6 Cosmetic Difference} accepts formatting, comments, or whitespace-only changes; and \textbf{MV7 Insufficient Evidence} is used only when the missing evidence is identified.
Two independent blinded verification passes label identical packages containing the pre-fix function, the tool output function, the reference function, their diffs, and local context as \emph{correct}, \emph{incorrect}, or \emph{uncertain}; each pass is conducted without tool identity or the other pass's label.
MV1--MV4 imply \emph{incorrect}, MV5--MV6 justify \emph{correct}, and a separate adjudication pass resolves every disagreement or uncertain label by reapplying the same checks to the shared evidence package and recording the rationale.
On the 315 retained function-level review items, the independent labels achieved Cohen's \(\kappa=0.5357\) with 76.5\% observed agreement; file-level success requires all reviewed functions in the file to pass, and commit-level success requires all files in the commit to pass.

\begin{table}[t]
  \centering
  \caption{Replication-dataset performance under the common static protocol}
  \label{tab:rq1-results}
  \resizebox{\columnwidth}{!}{%
  \begin{tabular}{l|cc|cc|cc}
  \toprule
  \multirow{2}{*}{\textbf{Tool}} & \multicolumn{2}{c|}{\textbf{Function}} & \multicolumn{2}{c|}{\textbf{File}} & \multicolumn{2}{c}{\textbf{Commit}} \\
  \cmidrule(lr){2-3} \cmidrule(lr){4-5} \cmidrule(lr){6-7}
   & EM & S-Succ & EM & S-Succ & EM & S-Succ \\
  \midrule
  FixMorph ($n$=1217/466/285) & 19.3\% & 19.6\% & 19.7\% & 20.4\% & 27.0\% & 28.1\% \\
  TSBPort ($n$=2020/917/600) & \textbf{67.7\%} & \textbf{68.8\%} & 70.9\% & 72.0\% & 74.5\% & 75.8\% \\
  Mystique$^\star$ ($n$=619/309/224) & 28.4\% & 30.0\% & 13.9\% & 14.9\% & 13.4\% & 14.7\% \\
  PPatHF$^\star$ ($n$=1440/570/367) & 22.3\% & 23.0\% & 21.9\% & 22.3\% & 29.4\% & 30.0\% \\
  PortGPT$^\star$ ($n$=2020/917/600) & \textbf{67.7\%} & 68.6\% & \textbf{73.6\%} & \textbf{75.0\%} & \textbf{78.5\%} & \textbf{80.5\%} \\
  \bottomrule
  \end{tabular}%
  }
  \parbox{\columnwidth}{\footnotesize Note: Static EM/S-Succ under the common protocol. $n$ gives the function/file/commit denominator for each tool; $^\star$ uses the subset aligned with the tool's released evaluation.}
\end{table}

\textbf{Non-exact breakdown.} At commit level, we analyze all \(EM=false\) entries by L2--L5, their conjunction, and the result after ManualVerify; entries without comparable tool output remain in the denominator and are treated as failures (Table~\ref{tab:non-exact-breakdown}).

\subsubsection{Evaluation Settings and Alignment}
\label{subsubsec:eval-scope}

Following PortGPT~\cite{portgpt2025}, we separate Localization (identifying where to apply the patch) from Transformation (generating the adapted patch code) to avoid conflating target identification with patch adaptation.

We adopt an End-to-End (E2E) evaluation protocol that jointly evaluates localization and transformation.
The denominator is the ground-truth (GT) baseline.
Any pipeline failure (missing outputs, parsing failures, patch-apply failures, or localization failures) counts as 0.
We report EM and S-Succ against GT.

To analyze the impact of available information, we define three information levels.
\textbf{I0} (repository-level): only the source commit and target repository are provided.
The tool must locate the target file and function.
\textbf{I1} (file-level): the target file path is given.
\textbf{I2} (function-level): the target function is given.

Because tools operate at different native execution granularities (Table~\ref{tab:tool-summary}), we run each tool in its native mode and align all outputs back to the same ground-truth coordinate space.
For LLM-based tools that support variable context (Mystique and PPatHF), we additionally scale the provided context (file-level vs.\ function-level) to analyze sensitivity to localization information.
For the primary stratified analyses, each tool is evaluated at its highest supported information level: I2 for Mystique, PPatHF, and PortGPT, and I1 for TSBPort.
We normalize all outputs back to ground-truth file/line coordinates, and a generated patch counts as a valid candidate only if it modifies the correct file and function scope. This normalization mirrors the hunk-to-function mapping used in ground-truth construction: tool outputs produced at hunk, function, or file granularity are projected onto the same commit$\to$file$\to$function coordinate system. Consequently, function-, file-, and commit-level results are aggregated against the same ground-truth hunk-derived denominator for every tool.

Table~\ref{tab:rq2-granularity} examines how E2E performance changes under I0--I2.
Using this framework, we evaluate the selected tools under the experimental settings described next.

\begin{table}[t]
  \centering
  \caption{Static E2E performance by porting type after ManualVerify}
  \label{tab:rq2-porting-type}
  \footnotesize
  \setlength{\tabcolsep}{3pt}
  \resizebox{\columnwidth}{!}{%
  \begin{tabular}{ll|ccc|ccc|ccc|ccc}
  \toprule
  \multirow{2}{*}{\textbf{Tool}} & \multirow{2}{*}{\textbf{Agg.}} & \multicolumn{3}{c|}{\textbf{Type-I}} & \multicolumn{3}{c|}{\textbf{Type-II}} & \multicolumn{3}{c|}{\textbf{Type-III}} & \multicolumn{3}{c}{\textbf{Type-IV}} \\
  \cmidrule(lr){3-5} \cmidrule(lr){6-8} \cmidrule(lr){9-11} \cmidrule(lr){12-14}
   & & $n$ & EM & S-Succ & $n$ & EM & S-Succ & $n$ & EM & S-Succ & $n$ & EM & S-Succ \\
  \midrule
  \multirow{3}{*}{TSBPort}  & func   & 437 & 34.1\% & 34.1\% & 1067 & 41.0\% & 41.3\% & 440 & 37.0\% & 37.3\% & 571 & 20.1\% & 20.5\% \\
                            & file   & 230 & 42.6\% & 42.6\% & 539 & 31.2\% & 31.7\% & 206 & 22.3\% & 23.8\% & 324 & 15.4\% & 15.4\% \\
                            & commit & 162 & 35.8\% & 35.8\% & 297 & 21.2\% & 21.2\% & 79 & 6.3\% & 7.6\% & 96 & 10.4\% & 10.4\% \\
  \midrule
  \multirow{3}{*}{Mystique} & func   & 437 & 39.1\% & 39.4\% & 1067 & 36.4\% & 36.5\% & 440 & 31.6\% & 31.6\% & 571 & 17.5\% & 17.5\% \\
                            & file   & 230 & 30.4\% & 30.9\% & 539 & 23.0\% & 23.2\% & 206 & 8.3\% & 8.3\% & 324 & 4.3\% & 4.3\% \\
                            & commit & 162 & 32.1\% & 32.7\% & 297 & 20.9\% & 21.2\% & 79 & 5.1\% & 5.1\% & 96 & 4.2\% & 4.2\% \\
  \midrule
  \multirow{3}{*}{PPatHF}   & func   & 437 & 32.7\% & 33.6\% & 1067 & 20.1\% & 20.1\% & 440 & 14.3\% & 14.8\% & 571 & 8.9\% & 9.1\% \\
                            & file   & 230 & 25.7\% & 26.5\% & 539 & 16.3\% & 16.3\% & 206 & 3.4\% & 3.9\% & 324 & 3.1\% & 3.1\% \\
                            & commit & 162 & 27.8\% & 29.0\% & 297 & 15.2\% & 15.2\% & 79 & 3.8\% & 3.8\% & 96 & 2.1\% & 2.1\% \\
  \midrule
  \multirow{3}{*}{PortGPT}  & func   & 437 & 87.0\% & 87.0\% & 1067 & 84.9\% & 84.9\% & 440 & 70.2\% & 72.0\% & 571 & 30.3\% & 31.2\% \\
                                      & file   & 230 & 88.3\% & 88.7\% & 539 & 82.2\% & 83.5\% & 206 & 61.2\% & 69.9\% & 324 & 32.1\% & 34.6\% \\
                                      & commit & 162 & 84.6\% & 85.2\% & 297 & 73.1\% & 74.4\% & 79 & 35.4\% & 49.4\% & 96 & 22.9\% & 24.0\% \\
  \bottomrule
  \end{tabular}%
  }
  \parbox{\columnwidth}{\footnotesize Note: Static E2E EM/S-Succ under the common protocol after ManualVerify. Type-I/II/III/IV denote identical, location-only, syntactic, and structural adaptation. $n$ is the denominator for the corresponding aggregation level.}
\end{table}

\subsection{Experimental Setup}
\label{subsec:setup}

All experiments were conducted on a server with 2$\times$ Intel Xeon Gold 6226R CPUs and 4$\times$ NVIDIA A100 (80GB) GPUs.
Each LLM-based tool was run with its originally recommended model and inference settings.
Mystique uses a locally deployed CodeLlama-13B-Instruct~\cite{codellama2023} with LoRA~\cite{lora2022} fine-tuning.
PPatHF uses a locally deployed StarCoder~\cite{starcoder2023} with LoRA fine-tuning (temperature 0.1).
PortGPT invokes GPT-4o~\cite{gpt4o2024} via the OpenAI API~\cite{openai_api} (temperature 0.5).
FixMorph and TSBPort are traditional program-analysis tools with no LLM component.
For FixMorph, following Mystique~\cite{mystique2024}, which evaluates FixMorph only on Linux repositories, we used the released package and also attempted the standalone release. The standalone release requires compilable \texttt{Pa}/\texttt{Pb}/\texttt{Pc} versions and build and configuration commands for each project to construct complete ASTs. Because validated configurations were available only for the original Linux kernel experiments, we evaluate FixMorph there in RQ1 and omit it from RQ2.
On the 634-case Evaluation Dataset, mean generation time is 4.0\,min/case for TSBPort, 2.2\,min/case for PortGPT (\$0.38/case), and 69.9\,s/request and 34.2\,s/request for local Mystique and PPatHF, respectively, with no API charge.
The primary eight-worker SWE-agent (GPT-4o) 45-case timestamp batch took 7.2\,min and cost \$21.04 (\$0.47/case).

\subsection{Research Questions}

We organize the paper around five research questions spanning tool performance, scenario-specific behavior, the gap between benchmark scores and real-world security effectiveness, root-cause explanation, and executable-feedback improvement:

\begin{itemize}[leftmargin=*,noitemsep]
    \rqitem{rq:stability}{Performance Stability}{How do existing tools perform under a common evaluation protocol compared with the results reported in their original papers?}
    \rqitem{rq:scenario}{Scenario Analysis}{How do tools perform across different backporting scenarios and complexity dimensions (porting type, cross scenario, information level, and patch scale)?}
    \rqitem{rq:gap}{Benchmark-to-Reality Gap}{Does reference agreement predict executable vulnerability remediation?}
    \rqitem{rq:failure}{Root-Cause Analysis}{What root causes explain the failure patterns and benchmark-to-reality mismatches observed above?}
    \rqitem{rq:improve}{Executable-Feedback Improvement}{Can executable feedback help improve the hardest cases identified above, especially Type-III/IV and cross-version patches?}
\end{itemize}

\section{Evaluation Results}
\label{sec:evaluation}

\begin{table}[t]
  \centering
  \caption{Static E2E performance by cross scenario after ManualVerify}
  \label{tab:rq2-cross-scenario}
  \resizebox{\columnwidth}{!}{%
  \begin{tabular}{ll|ccc|ccc|ccc}
  \toprule
  \multirow{2}{*}{\textbf{Tool}} & \multirow{2}{*}{\textbf{Agg.}} & \multicolumn{3}{c|}{\textbf{Cross-version}} & \multicolumn{3}{c|}{\textbf{Cross-branch}} & \multicolumn{3}{c}{\textbf{Cross-repository}} \\
  \cmidrule(lr){3-5} \cmidrule(lr){6-8} \cmidrule(lr){9-11}
   & & $n$ & EM & S-Succ & $n$ & EM & S-Succ & $n$ & EM & S-Succ \\
  \midrule
    \multirow{3}{*}{TSBPort}  & func   & 1260 & 34.4\% & 34.8\% & 908 & 36.3\% & 36.3\% & 347 & 29.1\% & 29.4\% \\
                              & file   & 558 & 22.2\% & 22.6\% & 547 & 34.4\% & 34.4\% & 194 & 25.8\% & 27.8\% \\
                              & commit & 233 & 12.4\% & 12.4\% & 283 & 27.2\% & 27.2\% & 118 & 25.4\% & 26.3\% \\
  \midrule
    \multirow{3}{*}{Mystique} & func   & 1260 & 33.0\% & 33.0\% & 908 & 31.8\% & 31.8\% & 347 & 26.8\% & 27.4\% \\
                              & file   & 558 & 18.3\% & 18.3\% & 547 & 16.3\% & 16.3\% & 194 & 17.5\% & 18.6\% \\
                              & commit & 233 & 18.9\% & 18.9\% & 283 & 20.1\% & 20.1\% & 118 & 17.8\% & 19.5\% \\
  \midrule
  \multirow{3}{*}{PPatHF}   & func   & 1260 & 12.8\% & 12.9\% & 908 & 20.4\% & 20.9\% & 347 & 36.0\% & 36.3\% \\
                              & file   & 558 & 9.3\% & 9.3\% & 547 & 12.1\% & 12.4\% & 194 & 23.7\% & 24.2\% \\
                            & commit & 233 & 10.3\% & 10.3\% & 283 & 14.1\% & 14.8\% & 118 & 26.3\% & 26.3\% \\
  \midrule
    \multirow{3}{*}{PortGPT}  & func   & 1260 & 69.1\% & 69.5\% & 908 & 74.1\% & 74.9\% & 347 & 64.6\% & 64.8\% \\
                                        & file   & 558 & 64.3\% & 67.7\% & 547 & 70.4\% & 72.2\% & 194 & 68.0\% & 70.6\% \\
                                        & commit & 233 & 59.2\% & 63.1\% & 283 & 68.9\% & 70.7\% & 118 & 60.2\% & 62.7\% \\
  \bottomrule
  \end{tabular}%
  }
    \parbox{\columnwidth}{\footnotesize Note: Static E2E EM/S-Succ under the common protocol after ManualVerify. Agg.\ = aggregation granularity; $n$ is the denominator for the corresponding aggregation level.}
\end{table}

\rqsubsection{rq:stability}{Performance Stability}{subsec:rq-stability}

Existing tools report performance on their own datasets with tool-specific success criteria, making comparison across tools difficult.
RQ1 therefore asks whether reported results remain stable after aligning the evaluation protocol while keeping tool usage conditions unchanged.
We evaluate all five tools on the 600-case Replication Dataset under the common evaluation metrics.
For fairness, each tool keeps its native input format and recommended setup (e.g., repository-level for FixMorph, function-level for Mystique), and we report outcomes at function-, file-, and commit-level aggregation.
As detailed in Section~\ref{subsubsec:data-sources}, tool-specific replication subsets are uneven (Mystique: 224/600, PPatHF: 367/600, FixMorph: 285/600, TSBPort/PortGPT: 600/600), so the per-tool denominator $n$ in Table~\ref{tab:rq1-results} differs accordingly.

Under the common protocol, commit-level performance separates into two groups.
PortGPT and TSBPort remain comparatively high, reaching 80.5\% and 75.8\% S-Succ respectively, although both are still below their originally reported results (89.2\%~\cite{portgpt2025} and 87.6\%~\cite{tsbport2023}).
The remaining tools lose much more: Mystique drops from 92.4\%~\cite{mystique2024} to 14.7\%, FixMorph from 75.1\%~\cite{fixmorph2021} to 28.1\%, and PPatHF from 42.3\%~\cite{ppathf2024} to 30.0\%.
These gaps may reflect incompatible native metrics (PDG matching, compilation-filtered equivalence, edit-distance thresholds), differences in original dataset difficulty, and the stricter func$\to$file$\to$commit ALL aggregation under the common protocol.

\finding{find:stability-gap}{Under the common protocol, all tools drop relative to their original reports; PortGPT and TSBPort remain high at commit level, while FixMorph and Mystique suffer the largest losses.}

\rqsubsection{rq:scenario}{Scenario Analysis}{subsec:rq-scenario}

We evaluate all tools on our Evaluation Dataset under the unified static evaluation scope defined above.
The official aligned denominator is therefore 634 commits / 1{,}299 files / 2{,}515 functions for the static results.
The Evaluation Dataset contains real-world cases, and we impose no target scenario ratios.
Its observed distribution describes the composition of this benchmark and may differ from the proportions across all backporting activity.
Unless otherwise noted, all tables report function-, file-, and commit-level results under the common evaluation protocol, using the same denominators and aggregation rules described above.

\begin{table}[t]
  \centering
  \caption{Macro-averaged commit-level S-Succ across RQ2 groups after ManualVerify}
  \label{tab:rq2-group-average}
  \footnotesize
  \setlength{\tabcolsep}{6pt}
  \begin{tabular}{l|cc}
  \toprule
  \textbf{Tool} & \textbf{By porting type} & \textbf{By cross scenario} \\
  \midrule
  TSBPort  & 18.8\% & 22.0\% \\
  Mystique & 15.8\% & 19.5\% \\
  PPatHF   & 12.5\% & 17.1\% \\
  PortGPT  & 58.2\% & 65.5\% \\
  \bottomrule
  \end{tabular}
  \parbox{\columnwidth}{\footnotesize Note: Each porting type in Table~\ref{tab:rq2-porting-type} or cross scenario in Table~\ref{tab:rq2-cross-scenario} contributes equally to the corresponding average.}
\end{table}

\subsubsection{Results by Porting Type}

Table~\ref{tab:rq2-porting-type} stratifies performance by porting type.

Performance generally decreases as adaptation requirements become more complex.
PortGPT remains the strongest tool across all four porting types, but its commit-level S-Succ still falls from 85.2\% on Type-I to 24.0\% on Type-IV.
The same pattern appears for the other tools at lower absolute rates: Type-IV reaches 10.4\% for TSBPort, 4.2\% for Mystique, and 2.1\% for PPatHF.
Thus, even when a tool remains comparatively strong on simpler or location-only ports, structural adaptation remains difficult under the common protocol.

  \finding{find:type-complexity}{Porting type is a major difficulty driver: performance drops sharply on Type-III/IV patches, and the best Type-IV commit-level S-Succ remains only 24.0\%.}

\subsubsection{Results by Cross Scenario}

Table~\ref{tab:rq2-cross-scenario} shows that scenario difficulty is tool-dependent under the common protocol after ManualVerify.
PortGPT remains strongest in all three scenarios, reaching 70.7\%, 63.1\%, and 62.7\% commit-level S-Succ on cross-branch, cross-version, and cross-repository cases, respectively.
The other tools remain much lower: their best scenario-level commit S-Succ ranges from 19.5\% to 27.2\%, and cross-version remains especially difficult for TSBPort and PPatHF.
This ordering is consistent with dataset characteristics (Figure~\ref{fig:eval-dataset}).
Cross-version cases have the largest share of complex adaptations (Type-III+IV = 40.3\%, versus 19.4\% and 22.0\%) and more evaluated files per commit (2.4 vs.\ 1.9 and 1.6).
Thus, scenario labels expose generalization differences, but the structural composition within each scenario determines how sharply each tool degrades.
The Evaluation Dataset was not collected under FixMorph's original Linux kernel setting.
Using FixMorph beyond that setting requires build and analysis configurations for each project, which were unavailable for the Evaluation Dataset. We therefore omit FixMorph from the remaining breakdowns.

Table~\ref{tab:rq2-group-average} complements the aggregate denominator with macro-averaged commit-level S-Succ across porting types and cross scenarios. These summaries reduce dependence on group sizes but are not context-independent tool rankings.

\finding{find:cross-scenario}{Cross-scenario generalization is tool-dependent: PortGPT leads across all scenarios, while cross-version remains difficult for the other tools.}

\subsubsection{Results by Information Level (I0--I2)}

Only PortGPT supports autonomous repository-level localization (I0); the other evaluated settings require supplied file paths (I1) or function code (I2). Additional localization information does not uniformly improve commit-level S-Succ: Mystique improves at I2, PPatHF remains essentially unchanged, and PortGPT stays stable across I0--I2 (Table~\ref{tab:rq2-granularity}).

A finer-grained analysis of PortGPT under I0 (phase-2 subset, $n{=}526$) shows that 86.3\% of cases produce analyzable localization outputs, but only 60.6\% hit the target function. Among the target-function hits, S-Succ reaches 79.3\%, indicating that substantial end-to-end loss occurs during localization.

\begin{table}[t]
  \centering
  \caption{Static E2E performance by information level}
  \label{tab:rq2-granularity}
  \resizebox{\columnwidth}{!}{%
  \begin{tabular}{l|cc|c|c}
  \toprule
  \multirow{2}{*}{\textbf{Tool}} & \multicolumn{2}{c|}{\textbf{I0 (repository)}} & \textbf{I1 (file)} & \textbf{I2 (function)} \\
  \cmidrule(lr){2-3} \cmidrule(lr){4-4} \cmidrule(lr){5-5}
    & \shortstack{Localization\\Coverage} & EM/S-Succ & EM/S-Succ & EM/S-Succ \\
  \midrule
    TSBPort   & --- & ---/--- & 21.5\%/21.6\% & ---/--- \\
    Mystique  & --- & ---/--- & 17.7\%/17.7\% & 19.2\%/19.6\% \\
    PPatHF    & --- & ---/--- & 15.3\%/15.6\% & 15.0\%/15.3\% \\
  PortGPT   & 86.5\% & 63.7\%/66.1\% & 62.8\%/65.8\% & 63.7\%/66.4\% \\
  \bottomrule
  \end{tabular}%
  }
    \parbox{\columnwidth}{\footnotesize Note: Commit-level static EM/S-Succ (\%). Localization coverage is reported only for PortGPT's autonomous target-file localization at I0; I1 supplies file paths and I2 supplies function context.}
\end{table}

\finding{find:localization}{Additional localization information is not uniformly beneficial: Mystique improves at I2, PPatHF remains essentially unchanged, and PortGPT stays stable across I0--I2, while its separate I0 stage analysis still shows substantial loss before function alignment.}

\subsubsection{Results by Patch Length and File Count}

Patch scale reinforces the same picture after ManualVerify. Once patches exceed 100 changed lines, only PortGPT remains above 40\% commit-level S-Succ in the 101--300-line bucket (44.2\%); in the 301--500-line bucket, PortGPT reaches 38.5\% while the other tools are at most 7.7\%, and all tools fall to 0.0\% beyond 500 changed lines. Multi-file changes are also brittle: PortGPT drops from 74.1\% on single-file commits to 26.3\% on commits touching more than five files, while TSBPort, Mystique, and PPatHF drop from 34.2\%, 30.2\%, and 24.9\% to 0.0\%. Detailed patch-scale tables and the replay script are included in the artifact.

\paragraph{Tolerance Sensitivity.}
We empirically evaluate \(\delta\in\{1,\ldots,5\}\) on both datasets and retain \(\delta=3\) as the smallest stable value.
Across these settings, commit-level S-Succ varies by at most 1.7\,pp per dataset, and tool ordering is unchanged.

\subsubsection{Non-Exact L2--L5 Breakdown}
\label{subsubsec:breakdown-sensitivity}

Table~\ref{tab:non-exact-breakdown} reports L2--L5 and ManualVerify results for all \(EM=false\) commits.

\begin{table}[t]
  \centering
  \caption{Commit-level L2--L5 breakdown for non-exact commits under $\delta$=3}
  \label{tab:non-exact-breakdown}
  \footnotesize
  \setlength{\tabcolsep}{2.5pt}
  \resizebox{\columnwidth}{!}{%
  \begin{tabular}{ll|c|cccc|c|c}
  \toprule
  \textbf{Dataset} & \textbf{Tool} & {\boldmath$n_{\neg\mathrm{EM}}$} & \textbf{L2} & \textbf{L3} & \textbf{L4} & \textbf{L5} & \textbf{L2--L5} & \textbf{+MV} \\
  \midrule
  \multirow{5}{*}{Replication}
    & FixMorph & 208 & 58.2\% & 25.0\% & 73.1\% & 24.5\% & 6.7\% & 1.4\% \\
    & TSBPort  & 153 & 20.9\% & 39.9\% & 35.3\% & 34.6\% & 16.3\% & 5.2\% \\
    & Mystique & 194 & 28.9\% & 24.2\% & 42.3\% & 23.2\% & 5.2\% & 1.5\% \\
    & PPatHF   & 259 & 15.8\% & 23.6\% & 25.1\% & 25.5\% & 2.3\% & 0.8\% \\
    & PortGPT  & 129 & 23.3\% & 41.9\% & 41.9\% & 34.1\% & 18.6\% & 9.3\% \\
  \midrule
  \multirow{4}{*}{Evaluation}
    & TSBPort  & 498 & 13.9\% & 25.7\% & 27.5\% & 13.5\% & 0.4\% & 0.2\% \\
    & Mystique & 522 & 2.3\% & 10.5\% & 3.3\% & 21.3\% & 0.2\% & 0.0\% \\
    & PPatHF   & 537 & 8.0\% & 7.4\% & 10.1\% & 9.3\% & 1.3\% & 0.4\% \\
    & PortGPT  & 236 & 17.4\% & 18.2\% & 19.1\% & 16.5\% & 12.3\% & 8.1\% \\
  \bottomrule
  \end{tabular}%
  }
  \parbox{\columnwidth}{\footnotesize Note: $n_{\neg\mathrm{EM}}$ is the number of non-exact commits; L2--L5 columns are independent pass rates, \textbf{L2--L5} is their conjunction, and \textbf{+MV} additionally applies ManualVerify. Replication rows use native replication settings; Evaluation rows use the common file-level setting where available (Mystique/PPatHF/PortGPT/TSBPort I1).}
\end{table}

The table shows that joint structural agreement is substantially stricter than any individual L2--L5 check: on the Evaluation Dataset, only 0.2\%--12.3\% of non-exact commits pass all four gates, and the result after ManualVerify is 0.0\%--8.1\%.

  \rqsubsection{rq:gap}{Benchmark-to-Reality Gap}{subsec:rq-gap}
  
The preceding results measure agreement with the reference patch, but such agreement does not establish executable remediation.
We therefore report \emph{POC Pass Rate} (PPR), for which the POC triggers on the vulnerable target but is blocked after applying the tool patch, and \emph{Test Pass Rate} (TPR), for which the configured developer test passes.
\emph{Full validation} requires both; PPR and TPR remain separate because either can succeed without the other.

From the 634-case Evaluation Dataset, we identify 179 candidates whose source or target reference patch modifies at least one test-related file.
We retain 45 satisfying the executable-evidence gate: a reproducibly buildable target, a runnable developer test, a behavioral POC that triggers on the vulnerable target, and a GT patch that applies and builds, blocks the POC, and passes the test.
Because these assets were iteratively curated rather than applied as independent sequential filters, we report the actual \(634\rightarrow179\rightarrow45\) transition.
All metrics retain 45 as the denominator; cases without a candidate patch and apply/build failures count as failures.
Because selection depends on the availability of executable evidence, results on these 45 cases are exploratory rather than representative of the full 634-case Evaluation Dataset~\cite{mu2018reproducing,hazimeh2020magma}.
To test whether executable failures are specific to PortGPT, we extend RQ3 to TSBPort, Mystique, and PPatHF from the RQ1/RQ2 baseline suite and include SWE-agent as a general coding agent.
All systems receive the common file-level (I1) input without build, test, or POC feedback.
For PortGPT, I1 supplies the target file while preserving iterative hunk adaptation through tool use; its autonomous localization from repository to file is evaluated separately under I0 in RQ2.
FixMorph is omitted because its released configuration cannot be aligned with this executable subset.

\begin{figure*}[t]
  \centering
  \begin{minipage}[t]{0.61\textwidth}
    \centering
    \includegraphics[width=\linewidth]{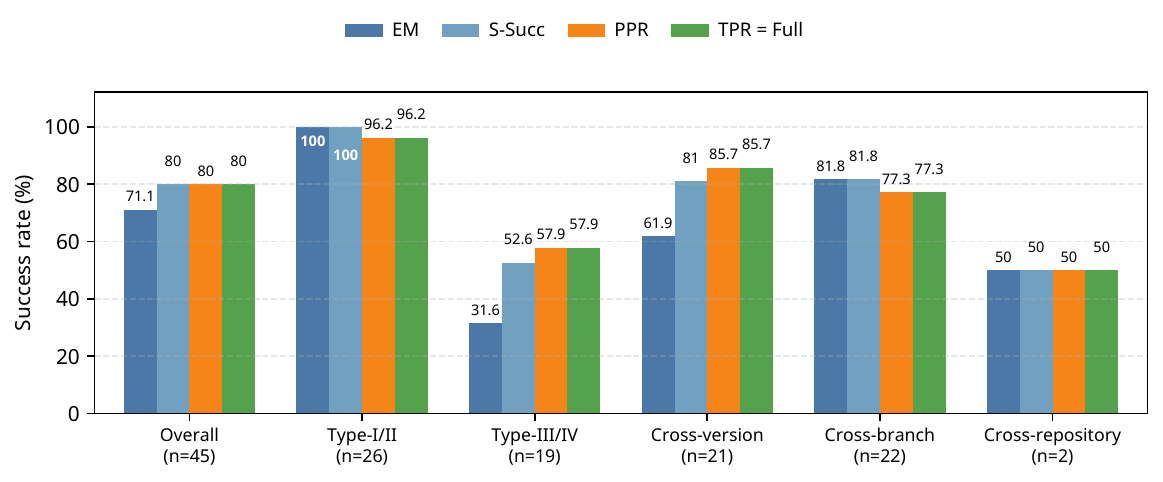}\\
    {\small (a) Grouped static and executable outcomes}
  \end{minipage}\hfill
  \begin{minipage}[t]{0.37\textwidth}
    \centering
    \includegraphics[width=\linewidth]{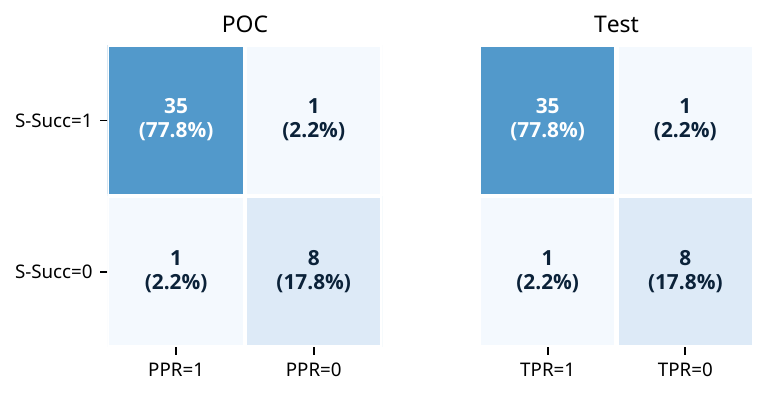}\\
    {\small (b) Case-level mismatch matrices}
  \end{minipage}
  \caption{PortGPT under the common I1 setting. Panel (a) reports EM and S-Succ separately across groups; TPR and Full coincide in every group. Panel (b) reports case-level agreement and mismatch; cases without a candidate patch count as failures.}
  \Description{A two-panel evaluation figure. Panel (a) is a grouped bar chart comparing EM, S-Succ, PPR, and TPR/Full across the overall set, porting types, and cross scenarios. Panel (b) contains two two-by-two matrices comparing S-Succ with PPR and TPR over the 45 executable cases, with 35 both-pass, 1 static-only, 1 executable-only, and 8 both-fail cases in each matrix.}
  \label{fig:rq3-gap-analysis}

\end{figure*}

Table~\ref{tab:crosstool-exec} compares executable outcomes across tools; Table~\ref{tab:gap-profile} and Figure~\ref{fig:rq3-gap-analysis} relate PortGPT's static and executable outcomes by group and case.
Although PortGPT's aggregate S-Succ and Full rates coincide at 36/45, two cases differ: one passes only S-Succ and one passes only executable validation.

\finding{find:gap}{SWE-agent outperforms three specialized baselines on Full validation but remains below PortGPT on the 45-case executable subset.}

\begin{table}[t]
  \centering
  \caption{Executable evaluation under the common I1 setting ($n{=}45$)}
  \label{tab:crosstool-exec}
  \footnotesize
  \begin{tabular}{l|ccc}
    \toprule
    \textbf{Tool} & \textbf{PPR} & \textbf{TPR} & \textbf{Full} \\
    \midrule
    TSBPort & 15.6\% (7/45) & 17.8\% (8/45) & 15.6\% (7/45) \\
    Mystique & 28.9\% (13/45) & 28.9\% (13/45) & 28.9\% (13/45) \\
    PPatHF & 28.9\% (13/45) & 28.9\% (13/45) & 28.9\% (13/45) \\
    SWE-agent (GPT-4o) & 37.8\% (17/45) & 42.2\% (19/45) & 37.8\% (17/45) \\
    PortGPT & \textbf{80.0\% (36/45)} & \textbf{80.0\% (36/45)} & \textbf{80.0\% (36/45)} \\
    \bottomrule
  \end{tabular}
  
\end{table}

\rqsubsection{rq:failure}{Root-Cause Analysis}{subsec:rq-failure}

We focus on the 175 Type-III/IV commits, which remain the most difficult porting-type group under the common protocol.
A stratified random sample of 120 failed tool--commit attempts with candidate and reference function evidence covers 92 unique commits and yields four primary root-cause categories:
\begin{itemize}[leftmargin=*,noitemsep]
    \item \textbf{F1: Missing target API awareness} (8.3\%). The generated patch references APIs or type definitions absent from the target. For example, several Fast-DDS cases retained \texttt{logError} instead of the target's \texttt{EPROSIMA\_LOG\_ERROR}.
    \item \textbf{F2: Cross-version semantic mismatch} (6.7\%). The same concept exists in both codebases but with different names or behavior; the generated patch uses the source form, which compiles but is functionally wrong. Unlike F1, compilation-based validation cannot catch such errors.
    \item \textbf{F3: Non-local dependency propagation failure} (39.2\%). The patch requires coordinated multi-function or multi-file edits, but the tool generates only partial changes. In the io\_uring case (Listing~\ref{lst:porting-example}), PortGPT produced only 2 header-level hunks with an incorrect bit position and omitted all 8 implementation-side hunks needed to thread the \texttt{iowait} parameter through a four-function call chain.
    \item \textbf{F4: Patch construction or localization failure} (45.0\%). The generated patch is empty at the required local site, malformed, placed in the wrong scope, or expanded with unrelated edits. This category captures failures where a narrower API or semantic diagnosis is blocked by the patch shape itself. One remaining sampled attempt (0.8\%) is a structural scoring anomaly and is not forced into F1--F4.
\end{itemize}

\finding{find:rootcause}{In the 120-attempt coding sample, patch construction or localization failure (F4, 45.0\%) and non-local dependency propagation failure (F3, 39.2\%) are the two most frequent categories, while missing target API awareness (F1, 8.3\%) and cross-version semantic mismatch (F2, 6.7\%) occur less frequently.}

\rqsubsection{rq:improve}{Executable-Feedback Improvement}{subsec:rq-improve}

\begin{table}[t]
  \centering
  \caption{PortGPT static and executable outcomes under the common I1 setting}
  \label{tab:gap-profile}
  \resizebox{\columnwidth}{!}{%
  \begin{tabular}{ll|c|cccc|cc}
    \toprule
    \textbf{Stratum} & \textbf{Sub-group} & $n$ &
      \textbf{EM} & \textbf{S-Succ} & \textbf{PPR} & \textbf{TPR} &
      \textbf{S-Succ$-$PPR} & \textbf{TPR$-$S-Succ} \\
    \midrule
    \multirow{4}{*}{Porting type}
      & Type-I   & 12 & 100.0\% & 100.0\% & 91.7\% & 91.7\% & 8.3 & $-8.3$ \\
      & Type-II  & 14 & 100.0\% & 100.0\% & 100.0\% & 100.0\%  & 0.0 & 0.0 \\
      & Type-III & 7  & 57.1\% & 57.1\% & 71.4\%  & 71.4\%  & $-14.3$ & 14.3 \\
      & Type-IV  & 12 & 16.7\% & 50.0\% & 50.0\%  & 50.0\%  & 0.0 & 0.0 \\
    \midrule
    \multirow{3}{*}{Cross scenario}
      & Cross-version    & 21 & 61.9\% & 81.0\% & 85.7\% & 85.7\% & $-4.7$ & 4.7 \\
      & Cross-branch     & 22 & 81.8\% & 81.8\% & 77.3\% & 77.3\% & 4.5 & $-4.5$ \\
      & Cross-repository & 2  & 50.0\% & 50.0\% & 50.0\%  & 50.0\%  & 0.0 & 0.0 \\
    \midrule
    Overall & --- & 45 & 71.1\% & 80.0\% & 80.0\% & 80.0\% & 0.0 & 0.0 \\
    \bottomrule
  \end{tabular}%
  }
  \parbox{\columnwidth}{\footnotesize Note: Commit-level results on the 45-case executable subset. Missing outputs and apply/build failures count as 0. PPR = exploit blocked; TPR = developer test passed; deltas are percentage points. Full requires both PPR and TPR and equals TPR in every row of this batch.}
  
\end{table}

The findings above motivate whether executable feedback can improve the hardest target adaptations identified in \rqref{rq:scenario} and analyzed in \rqref{rq:failure}, especially Type-III/IV and cross-version cases.
We select PortGPT because it achieves the strongest Full validation in RQ3 and its released workflow supports iterative refinement from execution feedback, whereas the other evaluated patch-porting tools do not provide the same mechanism.
Refinement covers only the 29-case \emph{hard scope} (Type-III/IV or cross-version) within the 45-case executable subset, so the observed gains are not generalized to other tools or to the full 634-case Evaluation Dataset.
Using corrected RQ3 as the frozen baseline, refinement is attempted only on non-Full cases with PortGPT I1 patches.
For each such patch, \texttt{git apply {-}{-}check} only checks whether the patch applies cleanly, after which we run the same apply, build, POC, and developer-test order as RQ3.
On failure, the refinement loop generates up to three SEARCH/REPLACE \emph{repair candidates} per round for at most four rounds, using the observed validation failure and patch context as feedback.
Within each round, repair candidates are evaluated in their fixed generation order against the current patch.
A candidate replaces the current patch only under an ordered, no-regression rule: it either reaches a later stage in the apply, build, POC, and developer-test sequence, or, at the same stage, preserves every previously passed check and passes a strict superset of developer tests.
If neither condition is met, the current patch is retained.
The loop stops immediately at Full validation; otherwise, it ends after four rounds.
Thus the loop tests whether executable feedback can mitigate F1--F3 without using GT-derived scoring during generation or selection.

\textbf{Answer to \rqref{rq:improve}.}
Table~\ref{tab:rq5-results} shows that refinement raises PortGPT's Full validation from 36/45 to 37/45 (+2.2\,pp).
Before refinement, the five eligible cases reached one Apply, three Build, and one POC as their highest validation stages; afterward, they reached one Build, three POC, and one Full.

\afterpage{%
\begin{table}[t]
  \centering
  \caption{RQ5 executable-feedback gains on the 45-case executable subset}
  \label{tab:rq5-results}
  \footnotesize
  \setlength{\tabcolsep}{2.0pt}
  \resizebox{\columnwidth}{!}{%
  \begin{tabular}{l|c|ccccc}
  \toprule
  \textbf{Slice} & $n$ &
  \shortstack{\textbf{Base}\\\textbf{Full}} &
  \textbf{Eligible} &
  \shortstack{\textbf{Stage}\\\textbf{+}} &
  \shortstack{\textbf{Final}\\\textbf{Full}} &
  \shortstack{\textbf{$\Delta$}\\\textbf{Full}} \\
  \midrule
  Type-III      & 7  & 71.4\% (5/7)  & 2 & 2 & 85.7\% (6/7)  & +1 \\
  Type-IV       & 12 & 50.0\% (6/12) & 2 & 1 & 50.0\% (6/12) & +0 \\
  Cross-version & 21 & 85.7\% (18/21) & 2 & 1 & 85.7\% (18/21) & +0 \\
  Hard scope    & 29 & 72.4\% (21/29) & 4 & 3 & 75.9\% (22/29) & +1 \\
  Overall       & 45 & 80.0\% (36/45) & 5 & 3 & 82.2\% (37/45) & +1 \\
  \bottomrule
  \end{tabular}
  }
  \parbox{\columnwidth}{\footnotesize
    Note: Base Full uses corrected RQ3; runner/materialization corrections are not RQ5 gains. Eligible counts corrected-RQ3 non-Full cases with PortGPT I1 patches. Stage+ may overlap with $\Delta$ Full. Rows may overlap.}
\end{table}
}


\section{Discussion}
\label{sec:discussion}

\subsection{Threats to Validity and Limitations}
\label{subsec:threats}

\textbf{Threats to validity.}
Root-cause categorization involves subjective judgment, and the reported proportions may depend on category boundaries.
EM and S-Succ compare against a reference patch and may under-credit structurally divergent correct fixes; ManualVerify only reviews non-exact functions that already pass L2--L5, so we report the non-exact breakdown and use PPR/TPR only as complementary executable evidence.
Executable findings and feedback gains are limited to 45 cases with reproducible POCs, developer tests, and build environments; small subgroup sizes prevent generalization to the complete benchmark~\cite{mu2018reproducing,hazimeh2020magma,openai2024swebenchverified}.

\textbf{Limitations.}
The results may not generalize beyond the C/C++ security patches drawn from over 50~projects~\cite{vpsurvey2024}.
One oracle-blocked and one incomplete case count as failures among the 45 cases but cannot be refined; FixMorph is excluded from the executable comparison because it did not run on newer kernels in our setup.
A temporal cutoff cannot rule out LLM training-data contamination.

\section{Related Work}
\label{sec:related}
\subsection{Automated Patch Backporting}

Automated patch backporting has been studied with program analysis, LLM-based prompting, and LLM agents~\cite{fixmorph2021,tsbport2023,mystique2024,ppathf2024,portgpt2025,llmport2025}. PatchWeave, SKYPORT, and MigGPT address narrower tasks: transplantation guided by exploits, backporting injection fixes, and migration of Linux patches maintained outside the main tree~\cite{patchweave2021,skyport2022,miggpt2025}. These tools use different datasets and scenarios, which makes direct comparison difficult~\cite{linuxporting2024,vulngap2025}. We did not evaluate LLMPort because its implementation was not public when we conducted the study. PatchFinder, ReBack, and PatchScope trace security fixes, recommend backports, or predict merge targets rather than transform patches~\cite{patchfinder2024,reback2024,patchscope2025}. Our benchmark evaluates transformation tools under a common static protocol and uses executable validation for difficult cases.

\subsection{Benchmarks for Software Engineering}

Software engineering benchmarks cover defects, vulnerable code, patch assessment, and secure code generation~\cite{defects4j2014,manybugs2015,bigvul2020,cvefixes2021,defects4c2025,cvebench2025,patcheval2025,vader2025,secureagentbench2025}. Defects4J and ManyBugs provide reproducible defects and test suites, whereas Big-Vul and CVEfixes organize vulnerability-fix data for mining tasks~\cite{defects4j2014,manybugs2015,bigvul2020,cvefixes2021}. BackportBench also uses execution outcomes, but it studies multilingual backporting at the package level rather than security patch backporting for C/C++~\cite{backportbench2025}. It does not report results by backporting scenario or study refinement with execution feedback, both of which our benchmark evaluates.

\subsection{Automated Program Repair}

Automated program repair (APR) creates a fix from tests or other fault evidence; patch backporting starts from a known fix and adapts it to a target that has changed~\cite{genprog2012,semfix2013,angelix2016,par2013,tbar2019}. Recent APR and patch validation systems use patch context and execution results to guide repair~\cite{chatrepair2024,repilot2023,giantrepair2025,reinfix2026,appatch2025,san2patch2025,repairagent2025,llmaprsurvey2025,patch2poc2026,patchvalidation2026,symradar2026}. Unlike APR, our task receives the source patch as repair intent and must preserve it across target-side syntactic, structural, and dependency changes.

\section{Conclusion}
\label{sec:conclusion}

Porting Benchmark shows that automated security patch backporting breaks mainly on structurally hard target adaptations: Type-III/IV and cross-version cases sharply reduce performance, static agreement can diverge from executable remediation, and executable feedback yields a small Full-validation gain plus stage-level progress.

\begin{acks}
Tencent Cloud Computing (Beijing) Co., Ltd. supported this research under Grant
No.~HX01202203011.
Additional support was provided by the National Research Foundation Singapore,
Prime Minister's Office, Singapore, and the Cyber Security Agency under the
National Cybersecurity R\&D Programme (NCRP25-P04-TAICeN) and its Campus for
Research Excellence and Technological Enterprise (CREATE) programme.
The China Scholarship Council (CSC) also supported this work under Grant
No.~202506960011.
\end{acks}

\section*{Data Availability Statement} Artifact metadata and available scripts are at \url{https://doi.org/10.5281/zenodo.21785770}.
\bibliographystyle{ACM-Reference-Format}
\bibliography{references}

\end{document}